\documentclass[twocolumn]{emulateapj}
\usepackage{epsfig}
\usepackage{amsmath}
\usepackage{graphicx}
\usepackage{natbib}
\usepackage{hyperref}
\shorttitle{Shape Evolution of Massive Early-type Galaxies}
\shortauthors{Chang et al.}
\begin{document}
\title{Shape Evolution of Massive Early-Type Galaxies: Confirmation of Increased
Disk Prevalence at $z>1$}
\author{Yu-Yen Chang\altaffilmark{1}}
\email{chang@mpia.de}
\author{Arjen van der Wel\altaffilmark{1}}
\author{Hans-Walter Rix\altaffilmark{1}}
\author{Stijn Wuyts\altaffilmark{2}}
\author{Stefano Zibetti\altaffilmark{3}}
\author{Balasubramanian Ramkumar\altaffilmark{1}}
\author{Bradford Holden\altaffilmark{4}}
\altaffiltext{1}{Max Planck Institute for Astronomy, Koenigstuhl 17, D-69117 Heidelberg, Germany}
\altaffiltext{2}{Max Planck Institute for Extraterrestrial Physics, Postfach 1312, Giessenbachstr., D-85741 Garching, Germany}
\altaffiltext{3}{INAF - Osservatorio Astrofisico di Arcetri, Largo Enrico Fermi 5, I-50125 Firenze, Italy}
\altaffiltext{4}{UCO/Lick Observatory, Department of Astronomy and Astrophysics, University of California, Santa Cruz, CA 95064, USA}
\begin{abstract}
We use high-resolution $K$-band VLT/HAWK-I imaging over 0.25 square degrees to study the structural evolution of massive early-type galaxies since $z\sim2$. Mass-selected samples, complete down to  $log(M/M_\odot)\sim10.7$ such that `typical' ($L^*$) galaxies are included at all redshifts, are drawn from pre-existing photometric redshift surveys. 
We then separated the samples into different redshift slices and classify them as late- or early-type galaxies on the basis of their specific star-formation rate. Axis-ratio measurements for the $\sim$400 early-type galaxies in the redshift range $0.6<z<1.8$ are accurate to 0.1 or better. The projected axis-ratio distributions are then compared with lower redshift samples. We find strong evidence for evolution of the population properties: early-type galaxies at $z>1$ are, on average, flatter than at $z<1$ and the median projected axis ratio at a fixed mass decreases with redshift. However, we also find that at all epochs $z \lesssim 2$ the very most massive early-type galaxies ($log(M/M_\odot)>11.3$) are the roundest, with a pronounced lack among them of galaxies that are flat in projection. Merging is a plausible mechanism that can explain both results: at all epochs merging is required for early-type galaxies to grow beyond $log(M/M_\odot)\sim11.3$, and all early types over time gradually and partially loose their disk-like characteristics.
\end{abstract}
\keywords{galaxies: evolution --- galaxies: formation --- galaxies: structure --- galaxies: elliptical and lenticular, cD --- cosmology: observations}
\section{Introduction}

In our theory of galaxy formation, the initial angular momentum, radiative energy loss of the gas, and some degree of angular momentum conservation result in gas settling into disks before most of the stars form, which makes for oblate, rotating stellar systems. Observationally, all but the most massive galaxies in the present-day universe have a disk-like structure and are rotating \citep{1980ApJ...236..351D,1989ARA&A..27..235K,1999ApJ...521...50M,2008MNRAS.390...93K,2009ApJ...693..617H,2011MNRAS.414..888E,2011MNRAS.412..727C}.

Nonetheless, even in disk-like, rotating galaxies, many stars reside in pressure-supported bulges. Photometric decompositions  indicate that in the present-day Universe 58$\pm$7\% of stars are in the spheroids and 42$\pm$7\% are in the disks \citep{2007MNRAS.379..841B}. Violent mergers are thought to scramble the orbits that originally were formed and lived in disks \citep{1977ARA&A..15..437T,1978MNRAS.183..341W,1985MNRAS.214...87J,1990ApJ...364L..33S,1993MNRAS.264..201K,2002NewA....7..155S}. This process dominates the evolution of the most massive galaxies, which do not show evidence for disks and are generally round and entirely supported by pressure instead of rotation (\citealt{1994ApJ...433..553J}; \citealt{2005ApJ...623..137V}; \citet[hereafter vdW09]{2009ApJ...706L.120V}; \citealt{2011MNRAS.412..684B}; \citealt{2011MNRAS.414..888E}; \citet[hereafter H12]{2012ApJ...749...96H}).

In this paper we use a low specific star formation as the definition
of early-type galaxy, motivated by the smooth appearance of the light
profiles of galaxies with little or no young stars or (star-forming)
gas. Early-type galaxies show a relatively abrupt change in their
structure as a function of galaxy mass. vdW09 and H12 show that galaxies with $log(M/M_\odot)<11$ have a broad projected axis-ratio distribution, indicative of a disk-like stellar body (with typical short-to-long intrinsic axis ratio about 1:3). In contrast, early-type galaxies with $log(M/M_\odot)>11$ rarely appear highly flattened, suggesting that their formation channel destroyed any pre-existing stellar disks, with `dry' merging between generally gas-poor progenitors as a very plausible mechanism.

Such `dry' mergers can reconcile the relatively late, and continuous, assembly of massive early types with their old, passively evolving stellar populations. In addition, the observed small radii and high densities of early-type galaxies at redshifts $z>1$ \citep[e.g.,][]{2008ApJ...677L...5V,2008ApJ...688...48V} can also be explained by `dry' merging \citep[e.g.,][]{2006MNRAS.370..902K,2009ApJ...698.1232V,2010MNRAS.401.1099H}. Minor mergers are plausibly the driver of size growth, given that
major mergers are less efficient in `puffing up' galaxies \citep[e.g.,][]{2009ApJ...697.1290B}, such that major merging would overproduce the
number of massive galaxies in the present-day universe \citep[e.g.,][]{2012arXiv1205.4058M}. Given the strong evidence for continuous evolution of the number of early-type galaxies \citep[e.g., ][]{2004ApJ...608..752B,2007ApJ...665..265F} and size evolution, one may expect that their structural properties also evolve, especially if merging is invoked as an evolutionary mechanism. To test this, H12 compared the axis-ratio distribution of a large sample of early-type galaxies at $z\sim0.7$ with that of the local population, but found no evidence for shape evolution for masses larger than $3\times10^{10} M_\odot$. However, there could be shape evolution at higher redshifts, as major merging occurred more frequently at early epochs \citep[e.g., ][]{2010ApJ...719..844R} and the early-type galaxy number was much lower at $z>1$ than it is today.

\citet[hearafter vdW11]{2011ApJ...730...38V} showed that a good portion of a small sample of $z\sim2$ early-type galaxies appeared flat in projection, indicative of a disk-like structure. \citet{2012ApJ...754L..24C} point out that existing, samples are too small to confirm or rule out evolution in the axis ratio distribution, but they attest, based on the S\'{e}rsic index distribution, that these galaxies have more disk-like structural properties than present-day early-type galaxies. Recently, \citet{2012arXiv1206.4322B} used bulge-disk decompositions of massive early-type galaxies at $z>1$ to show that many of them host pronounced disks. \citet{2012ApJ...745..179W} show a hint of an evolving axis ratio distribution of early-type galaxies out to $z\sim1.5$, but these authors did not explore this in detail.

In this paper we explore the (projected) shapes for a large sample ($\sim$400 objects) of early-type galaxies, selected to have masses $log(M/M_\odot)>10.7$ and low star formation rates ($sSFR<1/3t_H(z)$, see Section~2.5.), which were drawn from a wide, high-resolution, near-infrared ($K$-band) imaging mosaic from VLT/HAWK-I; we investigate whether early-type galaxies at $z>1$ show evolution in structure compared to present-day counterparts. Specifically, we will address the question whether $z>0.8$ early-type galaxies are more or less disk-like than at the present epoch, and whether early-type galaxies at those epochs also become rounder with increasing mass, as seen today.

The structure of this paper is as follows. In \textsection~2 we describe the data and select our sample of early-type galaxies. In \textsection~3 we analyze the projected axis-ratio distribution and its evolution since $z\sim2$. In \textsection~4 we summarize our conclusions.

In this paper, we use AB magnitudes and adopt the cosmological parameters ($\Omega_M$,$\Omega_\Lambda$,$h$)=(0.3,0.7,0.7).
\section{Data}

The first step is to compile a catalog with photometric redshifts and stellar masses of galaxies in the extended Chandra Deep Field South (ECDFS) from MUSYC \citep[Multiwavelength Survey by Yale-Chile,][]{2009ApJS..183..295T,2010ApJS..189..270C}. Then we use high-resolution VLT/HAWK-I K-band imaging available over essentially the full E-CDFS (ESO Program ID: 082.A-0890) to determine structural parameter (sizes, S\'{e}rsic indices and projected axis ratios) for these galaxies. The high fidelity of these measurements, verified through the comparison with results from HST imaging, allows us to select early-type galaxies up to $z=1.8$.

\subsection{Multi-Wavelength Data and SED Fitting}

MUSYC compiled observations in 32 bands, ranging from the UV to the near-infrared for the ECDFS, for which \citet{2010ApJS..189..270C} provide an optically selected catalog that we use here. We use the method and algorithms described by \citet{2011ApJ...742...96W} to
infer photometric redshifts, stellar masses and rest-frame colors.
Briefly, to estimate photometric redshifts ($z_{photo}$) we use EAzY \citep{2008ApJ...686.1503B}, and to estimate stellar masses, star-formation rates, and rest-frame colors we use FAST \citep{2009ApJ...700..221K}. We only include objects with significant detection of $J$, $H$ and $K$-band imaging, and reject stars by choosing only objects with $J-K>0.05$. We adopt the \citet{2003MNRAS.344.1000B} model, and a \citet{2003PASP..115..763C} stellar Initial Mass Function. A range of ages, star formation histories and extinction parameters is explored. This parent catalog contains 19642 objects.

H12 independently determined the stellar masses of early-type galaxies in the ECDFS in the redshift range $0.6<z<0.8$. Those mass estimates are designed to match the stellar mass estimates of present-day early-type galaxies. Since we aim to do the same we add 0.1 dex to all our stellar mass estimates to correct for the median difference between the galaxies that are included both in our parent sample and the H12 sample. This correction is likely incorrect for star-forming galaxies, but those are not considered in this work. A full investigation of the absolute mass scale for $z>1$ is beyond the scope of this paper.

Star formation rates (SFRs) were derived following the procedures outlined in \citet{2011ApJ...738..106W}. Briefly, the unobscured SFR traced by the $UV$ was added to the dust-reemitted SFR inferred from FIDEL 24 $\micron$ photometry \citep{2009A&A...496...57M} for 24 $\micron$-detected sources, and for sources without 24 $\micron$ detection a dust-corrected SFR was derived from stellar population modeling of the U-to-8 $\micron$ SED.
 
\subsection{High-Resolution, Near-Infrared VLT/HAWK-I imaging}
 
\begin{figure*}
\centering
\includegraphics[width=1.0\textwidth]{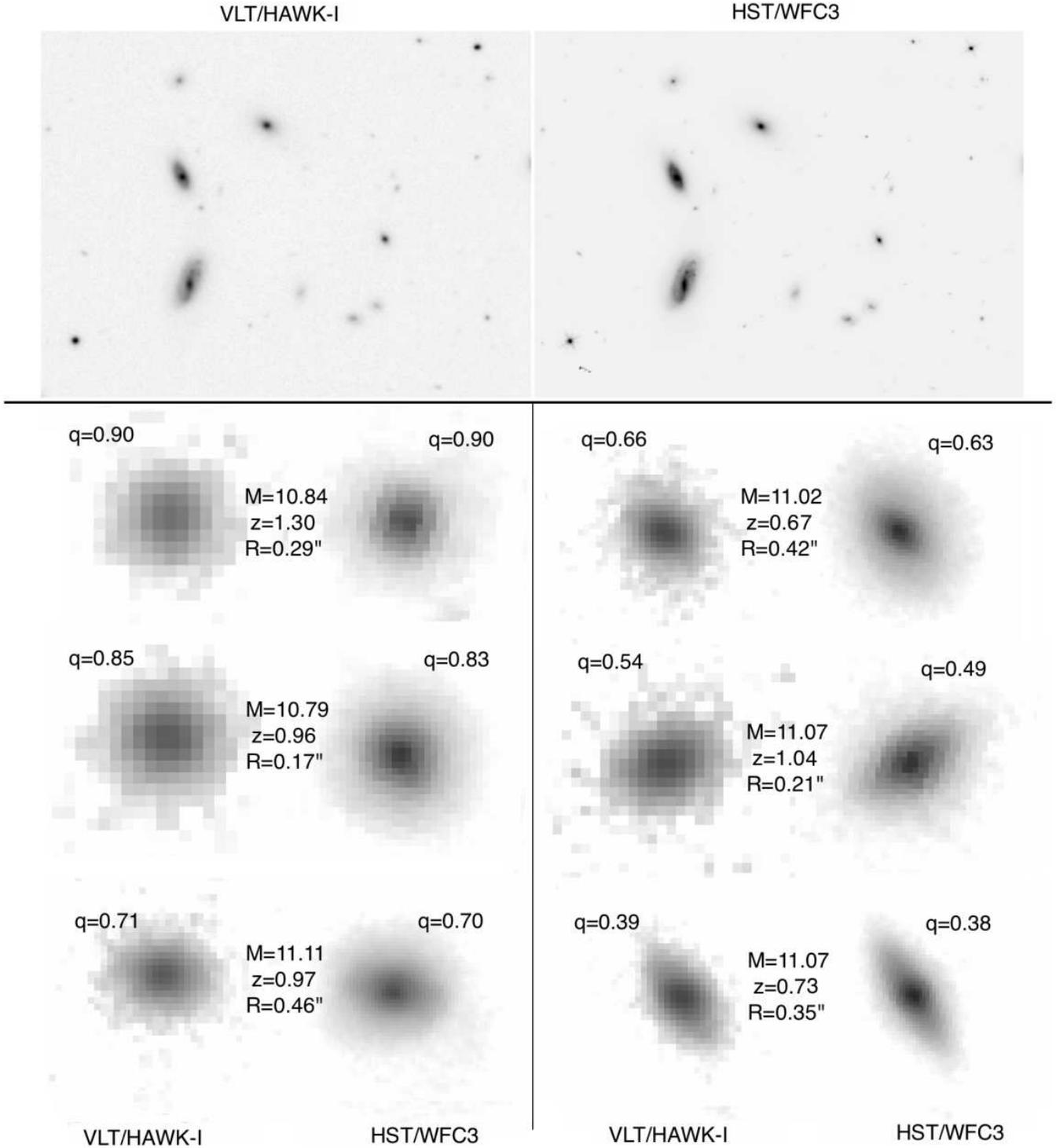}
\caption{Comparison between VLT/HAWK-I imaging (top left) used for the analysis in this paper HST/WFC3 imaging (top right; Early Release Science data from \citealt{2011ApJS..193...27W}) of the same galaxies. The bottom panels show zoomed-in versions of the images of the HAWK-I and WFC3 images of 6 early-type galaxies in the redshift, stellar mass range of interest and circularized half-light radii of HAWK-I, as indicated. Each panel is about 2'' on the side. These examples indicate that the resolution and depth of the HAWK-I imaging are sufficient to infer the projected axis ratio $q$, consistent with the values obtained from the WFC3 images.}
\label{vlt_hst_image}
\end{figure*}

High-resolution $K$-band imaging from VLT/HAWK-I, is central to our study to provide the structural parameters. We have obtained 1-hour exposures for each of 16 adjacent tiles in a 30'$\times$30' mosaic that covers the full ECDFS \citep{2009ApJS..183..295T}, which is coincident with the HST/ACS coverage from GEMS \citep[Galaxy Evolution from Morphologies and SEDs, ][]{2004ApJS..152..163R}. Observation and reduction of the HAWK-I images have been performed by S.~Z.~using a customized pipeline based on the original version distributed by ESO. Most notably we implemented improved recipes for the construction of the master flat field and for the frame coaddition, which properly take into account object masks and variance maps. The new effective mask implementation in particular eliminates the effects of background over-subtraction which is often seen in correspondence of (bright) sources. The 5-$\sigma$ point source limit is $K$(AB)=24.3, and the Point Spread Function (PSF) has a FWHM of 0.5'' or smaller across the field. At that resolution, this ground-based $K$-band imaging allows us to quantify the rest-frame optical structural properties of $z>1$ galaxies with spatial resolution that differs by no more than a factor of $\sim2$ from that obtained with HST/WFC3 in the $H$-band (see Figure~\ref{vlt_hst_image}). The image quality of our data is substantially better than the $\sim$0.6-0.7'' seeing data used by \citet{2012ApJ...745..179W}.

\subsection{Galaxy Structural Parameters}

\begin{deluxetable}{c|cc|ccc}
\tablewidth{0pc}
\tabletypesize{\scriptsize}
\tablecaption{Massive early-type galaxies: $0.6<z<1.8$, $log(M/M_{\odot})>10.7$ and $sSFR<1/3t_H(z)$. ID is the same as \citet{2010ApJS..189..270C}. Redshift ($z_{phot}$) and stellar mass ($M$) are from SED fitting. $K$-band magnitude($K$), effective radius ($R_e$) and projected axis ratio ($q_{proj}$) are from the {\tt GALFIT} results.}
\tablehead{
\colhead{ID} & \colhead{$z_{phot}$} & \colhead{M [$logM_\odot$]} & 
\colhead{$K [mag]$} & \colhead{$R_e$}  & \colhead{$q_{proj}$}
}
\startdata
 4032 &  1.15 & 10.72 & 20.68 &  0.25 &  0.49 \\
 4213 &  1.11 & 10.73 & 20.25 &  0.35 &  0.83 \\
 4619 &  1.00 & 10.87 & 21.72 &  0.09 &  0.87 \\
 4844 &  1.59 & 10.94 & 20.94 &  0.35 &  0.56 \\
 5375 &  1.13 & 11.33 & 19.46 &  0.44 &  0.96 \\
 . & . & . & . & . & . \\
 . & . & . & . & . & . \\
 . & . & . & . & . & . \\
\enddata
\label{gal}
\end{deluxetable}

We use {\tt GALAPAGOS} \citep[Galaxy Analysis over Large Areas: Parameter Assessment by GALFITting Objects from SExtractor, ][]{2012MNRAS.422..449B} to separately process each of the 16 HAWK-I tiles.  Here we briefly describe the process as relevant for the present study. For a full description, see \citet{2012MNRAS.422..449B}.~{\tt GALAPAGOS} first constructs a catalog with {\tt SExtractor} \citep[Software for source extraction, ][]{1996A&AS..117..393B}.  We choose the {\tt SExtractor} detection parameters such that the catalog is only complete down to K(AB)=23: as we will demonstrate below this is well beyond the limit down to which structural parameters can accurately be determined. {\tt GALAPAGOS} then creates image and noise cutouts for each object, including neighboring objects as necessary.  The noise map is obtained from the variance map that was produced for each of the HAWK-I tiles. The background is estimated for each object by identifying a set of sky pixels that are not influenced by any of the objects in the catalog.

For each of the 16 tiles a single star is taken as the PSF, chosen among the 5 brightest, isolated stars in each tile. This choice is made after subtracting a flux-scaled version of each of those 5 stars from $\sim$25 stars in a tile and examining the residuals from this fit.  The star that produces the cleanest residuals is selected as the PSF for that tile.

Then {\tt GALFIT} \citep[v3.0.3, ][]{2002AJ....124..266P, 2010AJ....139.2097P} is called to perform the actual measurement of the structural parameters.  A single S\'{e}rsic profile is fit to each target object. Neighboring objects are either masked or fit simultaneously. The free parameters in the fit are position, magnitude ($m$), effective radius as measured along the major axis ($R_e$), S\'{e}rsic index ($n$), axis ratio ($q$), and position angle.  The input values of these parameters are taken from the {\tt SExtractor} catalog (with the exception of $n$, for which 2.5 is adopted).

\subsection{Precision and Accuracy of Axis-Ratio Measurements}

\begin{figure}
\centering
\includegraphics[width=1.0\columnwidth]{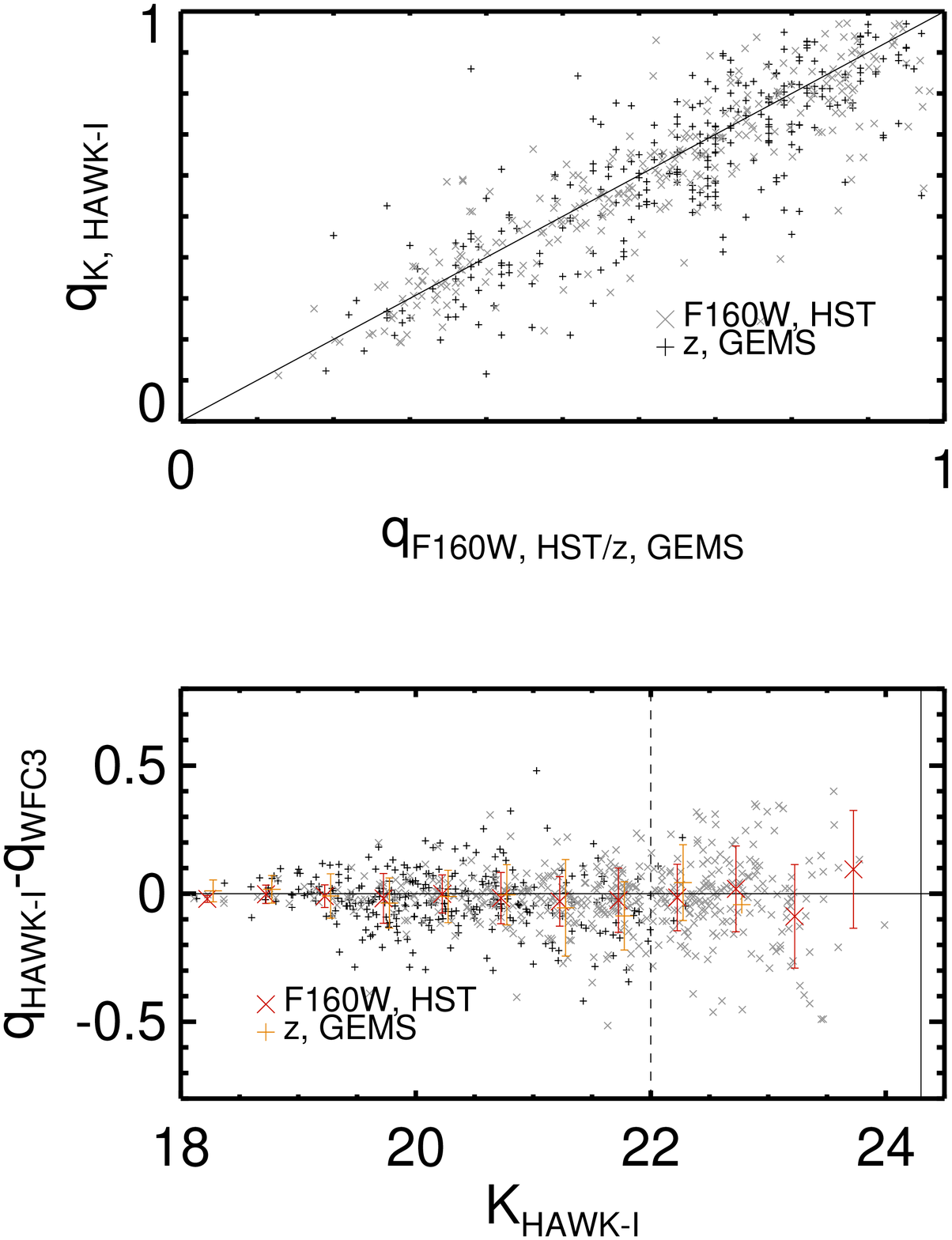}
\caption{Comparison of axis-ratio measurements from HAWK-I $K$-band and from HST imaging. The crosses represent the comparison with measurements from WFC3 F160W imaging for a HAWK-I $K$-band selected sample (regardless of galaxy type and redshift). The plus signs represent the comparison with measurements from HST/ACS F850LP of the H12 sample, which includes early-type galaxies in the redshift range $0.6<z<0.8$. The top panel directly compares the VLT and HST axis-ratio measurements; the bottom panel shows the difference between the VLT and HST measurements as a function of HAWK-I $K$-band magnitude. The red and orange bars represent the median and
standard deviation for a series of magnitude bins. The standard deviation represents the measurement uncertainty in the VLT-inferred axis ratio, assuming the HST-based value as `truth'. For galaxies, regardless of type, brighter than $K=22$ the uncertainty is smaller than or equal to 0.1, and we adopt this as the magnitude limit for our study. Interestingly, the accuracy of the axis-ratio measurement is good down to at least $K\sim24$: systematic effects in the axis-ratio measurements are small compared to the random uncertainty.}
\label{q_error}
\end{figure}

\begin{figure}
\centering
\includegraphics[width=1.0\columnwidth]{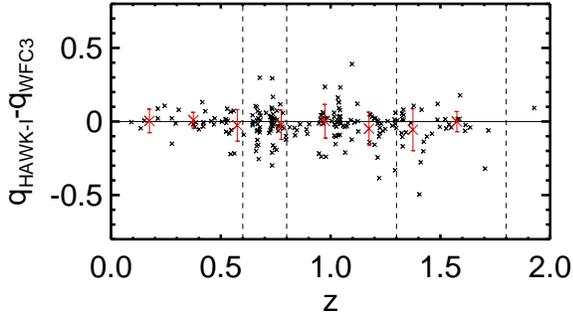}
\caption{Comparison of axis-ratio measurements from VLT/HAWK-I and HST/WFC3 imaging as a function of redshift. The difference in axis ratio measurements, $q_{HAWK-I}-q_{WFC3}$ has been multiplied by the total $K$ band flux in units of the K band flux for a $K=21$ object, such that variations in magnitude and their effect on the measurement uncertainty are accounted for. For galaxies in a narrow magnitude range (here, $20<K< 22$), there is no systematic redshift dependence: the uncertainty for galaxies at $0.5<z<1.0$ is comparable to the uncertainty for galaxies at $z>1$. The trends shown in Figure~\ref{q_error} and this figure imply that the accuracy of our axis ratio measurements is 10\% of better for all galaxies in our sample.}
\label{q_error_z}
\end{figure}

Our ability to determine axis ratios for distant galaxies is the limiting factor in our study.  To establish the precision and accuracy of our measurements we compare the axis ratios inferred from the HAWK-I imaging with those from GEMS\citep{2007ApJS..172..615H} for galaxies in the redshift range $0.6<z<0.8$ (the sample from H12). The wavelength difference between GEMS (z-band) and HAWK-I ($K$-band) could cause intrinsic differences between axis-ratio measurements that cannot be attributed to measurement errors.  For this reason, we also compare with the axis ratio estimates from objects in our sample that are contained with HST/WFC3 F160W imaging from ERS \citep{2011ApJS..193...27W}. {\tt GALAPAGOS} is deployed in a similar fashion as described above -- full details will be provided by van der Wel (2012, in prep.). Through the comparison with HST/ACS and HST/WFC3 (see Figure~\ref{q_error}) we find that the precision of our HAWK-I axis ratio estimates is better than 10\% for galaxies brighter than K(AB)=22. In addition, we see no systematic difference between HST and VLT measurements shown in Figure~\ref{q_error}. The differences in the median are -0.019 and -0.025, respectively. Our axis-ratio measurements remain accurate over the entire magnitude range of our sample.  In Figure~\ref{q_error_z} we show that, in addition, the accuracy of our axis ratio measurements does not depend on redshift.

\subsection{Sample Selection}

\begin{figure}
\centering
\includegraphics[width=1.0\columnwidth]{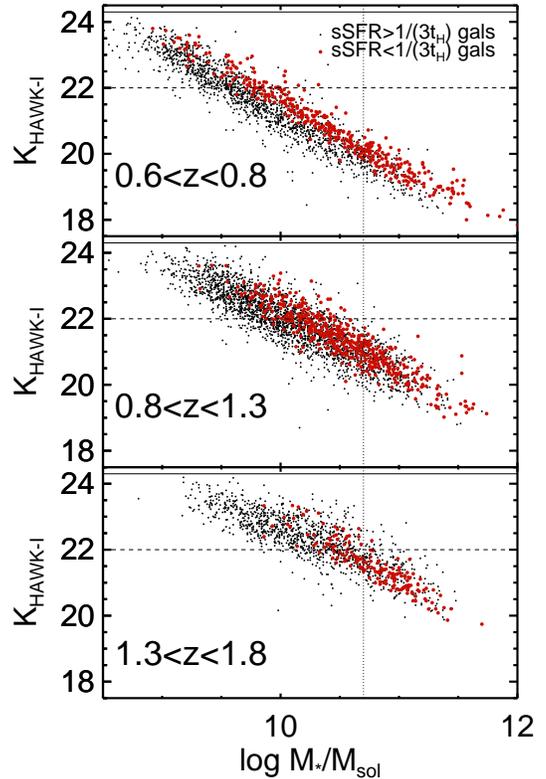}
\caption{HAWK-I $K$-band magnitude versus stellar mass in three redshift bins. Early-type galaxies, selected by their low sSFR (see Section 2.5.), are indicated by red symbols. The limiting factor in selecting our sample is set by the axis ratio estimate precision (see Figure~\ref{q_error}), which sets our magnitude limit to
$K=22$ (the horizontal lines). The implication is that in our highest redshift bin we are complete down to a mass limit of $log(M/M_\odot)\sim10.7$ (the vertical lines).}
\label{completeness}
\end{figure}

We match the MUSYC-based catalog (containing photometric redshifts and stellar population properties) with the HAWK-I based catalog (containing $K$-band magnitudes and structural parameters) by searching within apertures with radius 1''. In Figure~\ref{completeness} we show the distribution of $K$-band magnitude as a function of stellar mass in three redshift bins. Given the magnitude limit of $K=22$ that we adopted to ensure precise axis-ratio measurements (see above), we find that our catalog is complete down to $log(M/M_\odot)=10.7$ for all redshifts $z<1.8$. 

We select galaxies with stellar masses $log(M/M_\odot)>10.7$ and redshifts $0.6<z<1.8$, using spectroscopic redshifts when available, and, otherwise, photometric redshifts.
We select as early-type galaxies those with specific star-formation rates
($sSFR = SFR / M_{\odot}$) smaller than $1/3t_H(z)$, where $t_H(z)$ is the Hubble time
at the photometric redshift of the galaxy. This is similar to the strategy adopted by H12, who also
selected their samples by star-formation activity, rather than
morphological appearance.  

Visual morphological classifications are
not possible for the $z>1$ galaxies in our sample as they are too
faint. Automated classifiers based on, for example, concentration or
S\'{e}rsic index have the problem that, by definition, they will select
against disk-like objects.  This motivates our choice to select
early-type galaxies by their star-formation activity.  This is
justified further by the well-established agreement between
morphological appearance and star formation activity at all redshifts
$z<1$ \citep[e.g., ][]{2004ApJ...608..752B}, and the correlation between structure
and star formation activity that is observed to persist out to at
least $z\sim2$ \citep[e.g., ][]{2011ApJ...735L..22S,2012arXiv1208.0341P}.

\begin{figure}
\centering
\includegraphics[width=1.0\columnwidth]{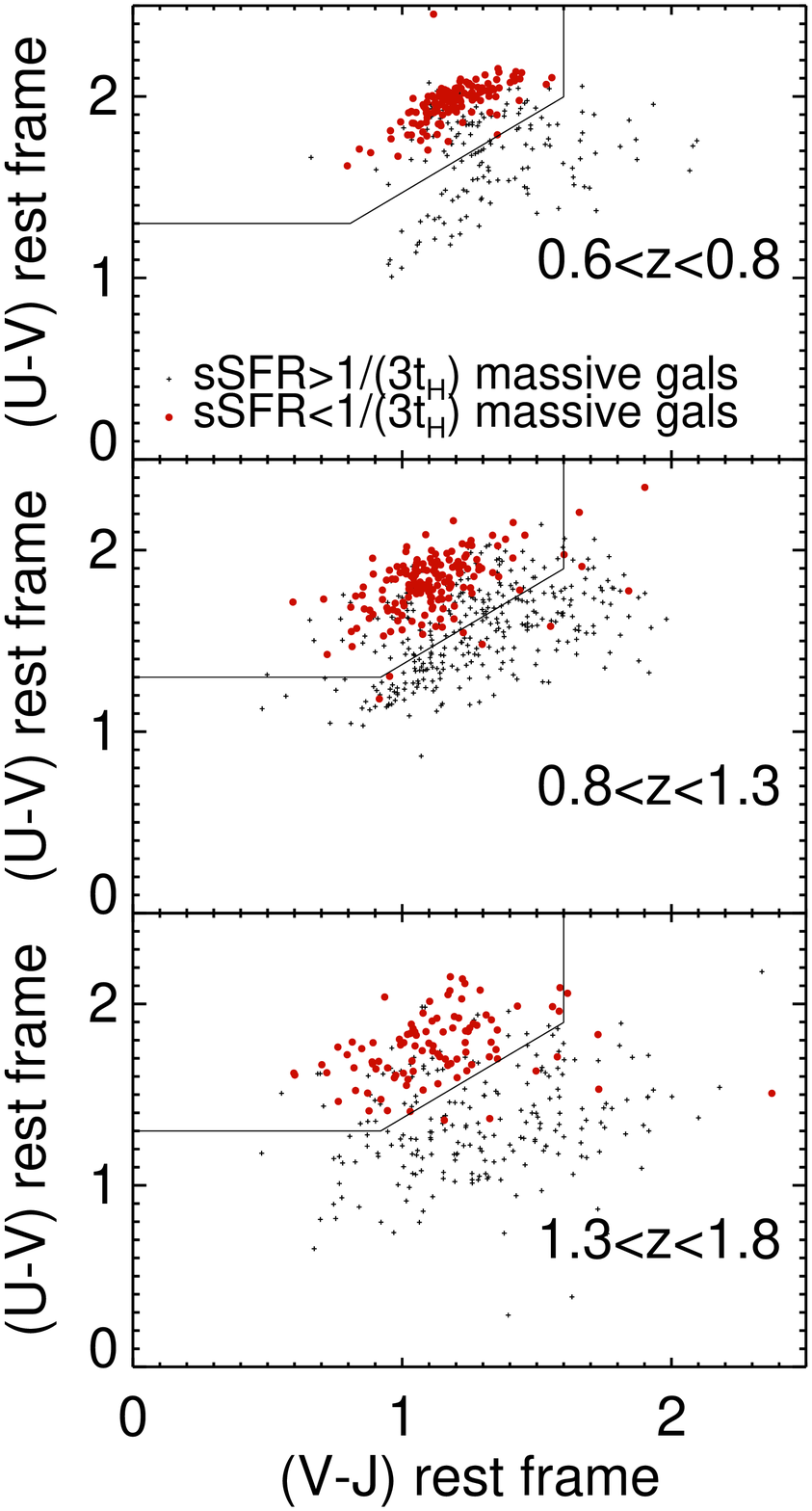}
\caption{Rest-frame $U$-$V$ color versus rest-frame $V$-$J$ color, a diagnostic diagram to distinguish passive and star-forming galaxies. The early-type galaxies in our sample, selected by having SED-based sSFR(=SFR/$M$)$<1/3t_H(z)$, are indicated by red symbols. These are essentially always located in the `passive box' as defined by \citet{2009ApJ...691.1879W}as indicated by the polygons. That our selection criteria are conservative is indicated by the presence of a substantial number of objects with $U$-$V$ and $V$-$J$ colors consistent with those of passive galaxies but with sSFR larger than our limit.}
\label{uvj}
\end{figure}

In Figure~\ref{uvj} we show that our selection technique is compatible with
the rest-frame $UVJ$ selection technique that is often adopted to
identify passive, early-type galaxies \cite[e.g, ][]{2007ApJ...655...51W,2009ApJ...691.1879W}.
Almost all of our low sSFR galaxies would also
be identified as early-type galaxies by selecting them in this $UVJ$ diagram. 
There is a substantial number of galaxies in the `passive' $UVJ$
color-color box that are star forming according to our direct star
formation estimates fits.  This number is far larger than the number
of galaxies with low star formation rates that are located outside the
`passive' $UVJ$ color box.  While we include the latter in our
subsequent analysis, we note that this does not affect our results.
Samples selected by $UVJ$, star-formation rate or a combination 
all have have the same median axis ratio within 0.05.

\begin{figure}
\centering
\includegraphics[width=1.0\columnwidth]{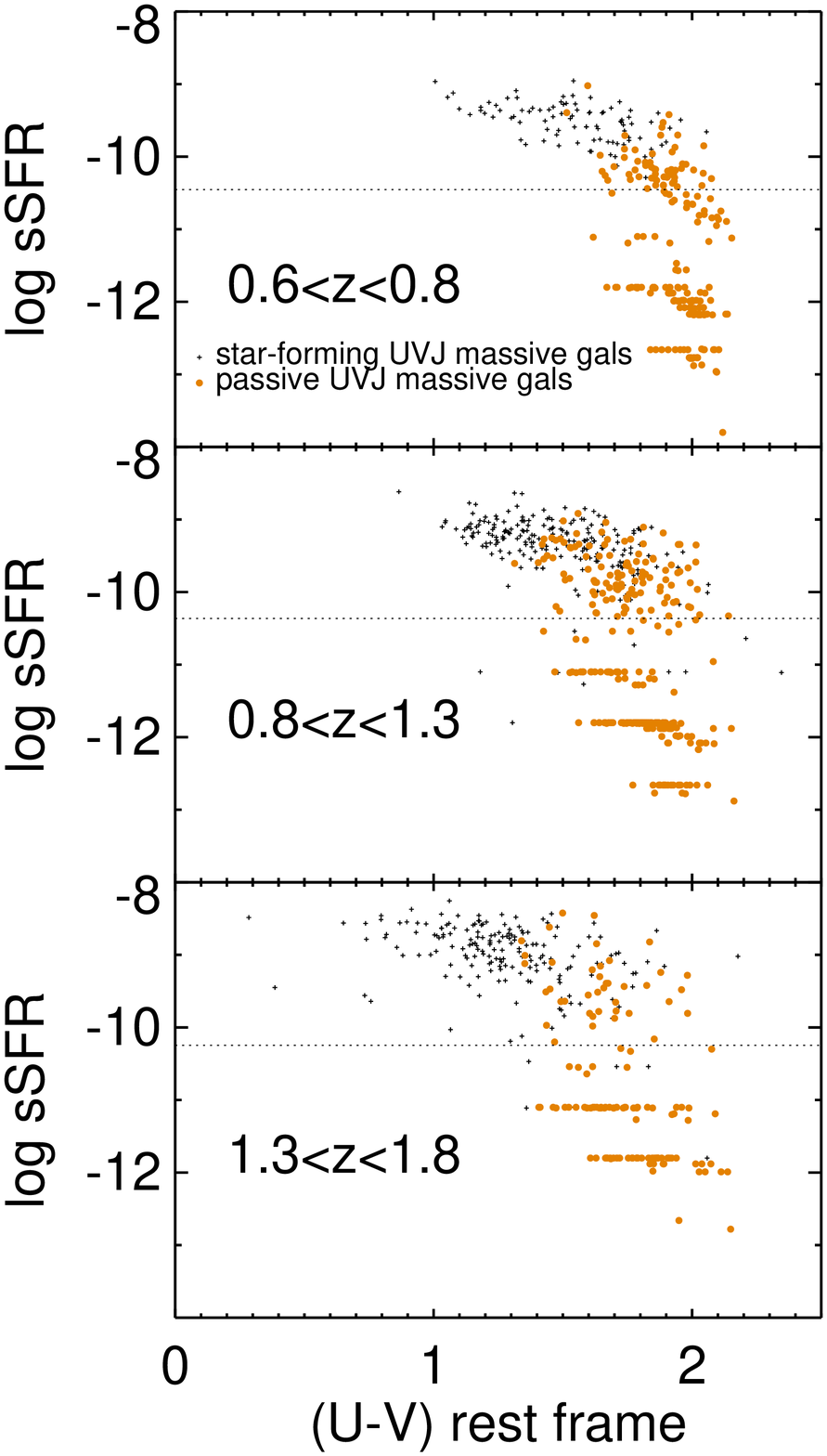}
\caption{The correlation between rest-frame $U-V$ color and sSFR for massive galaxies in the three redshift bins. The red symbols represent galaxies in the `passive' UVJ color-color box (see Figure~\ref{uvj}); the black symbols represent star-forming galaxies. Our selection criterion for early-type galaxies is based on a sSFR cut ($sSFR<1/3t_H(z)$), here illustrated by the horizontal, dotted lines which correspond to the sSFR cuts in the center of the redshift bins. For all redshift bins the two types of galaxies are cleanly separated; moving the sSFR cut by a modest amount does not strongly affect our selection of early-type galaxies. Note that aliasing occurs for low-sSFR objects as a result of template choices in the SED fitting.}
\label{uv_ssfr}
\end{figure}

In Figure~\ref{uv_ssfr} we show the correlation between rest-frame $U-V$ color and
sSFR.  Even in our highest redshift bin the populations of star-forming
and passive galaxies separate cleanly, which implies little 
cross-contamination between the two types of galaxies.

Finally, we visually inspected all {\tt GALFIT} fitting results, rejecting
10 objects with corrupted fits.  The final sample consists of 394
galaxies (see Table~\ref{median_mean}) in the redshift range $0.6<z<1.8$,
$\log(M/M_{\odot})>=10.7$, and $sSFR<1/3t_H(z)$. There are 134, 163 and 97
galaxies in the redshift bins, $0.6<z< 0.8$, $0.8<z<1.3$ and $1.3<z<1.8$, respectively.

\begin{deluxetable*}{c|ccccc}
\tablewidth{0pc}
\tabletypesize{\scriptsize}
\tablecaption{Statistical properties of $q_{proj}$} 
\tablehead{
\colhead{$log(M/M_\odot)>10.7$} & \colhead{number} & \colhead{median} & \colhead{mean} & \colhead{stddev} & \colhead{median of $z$} 
}
\startdata
$1.53<z<2.31$ (vdW11) & 14 & 0.665 & 0.681 & 0.186 & 1.69\\
$1.3<z<1.8$  & 97  & 0.634 & 0.612  & 0.200 & 1.38\\
$0.8<z<1.3$  & 163  & 0.646 & 0.619  & 0.213 & 1.09\\
$0.6<z<0.8$  & 134 & 0.734 & 0.662  & 0.222 & 0.69\\
$0.6<z<0.8$ (H12) & 533   & 0.710 & 0.692   & 0.179 &0.69\\
$0.04<z<0.08$ (vdW09) & 18316 & 0.719 & 0.692 & 0.181 &0.06
\enddata
\label{median_mean}
\end{deluxetable*}


\section{Structural Evolution of Massive Early-Type Galaxies}

We now examine two aspects of the structural evolution of early-type galaxies. First, we address the question whether the most massive galaxies are intrinsically round at redshifts $1<z<2$, as is observed at $z<1$ by vdW09 and H12. Second, we address the question whether, in general, early-type galaxies are flatter or rounder (e.g., disk-like or bulge-like) at $z>1$ than at $z<1$; H12 showed that there is little evolution up to $z<1$, but vdW11 presented tentative evidence for a higher incidence of disk-like early-type galaxies at $z\sim2$.

\subsection{The Mass-Dependence of Early-Type Galaxy Shapes up to $z\sim2$} 

\begin{figure}
\centering
\includegraphics[width=1.0\columnwidth]{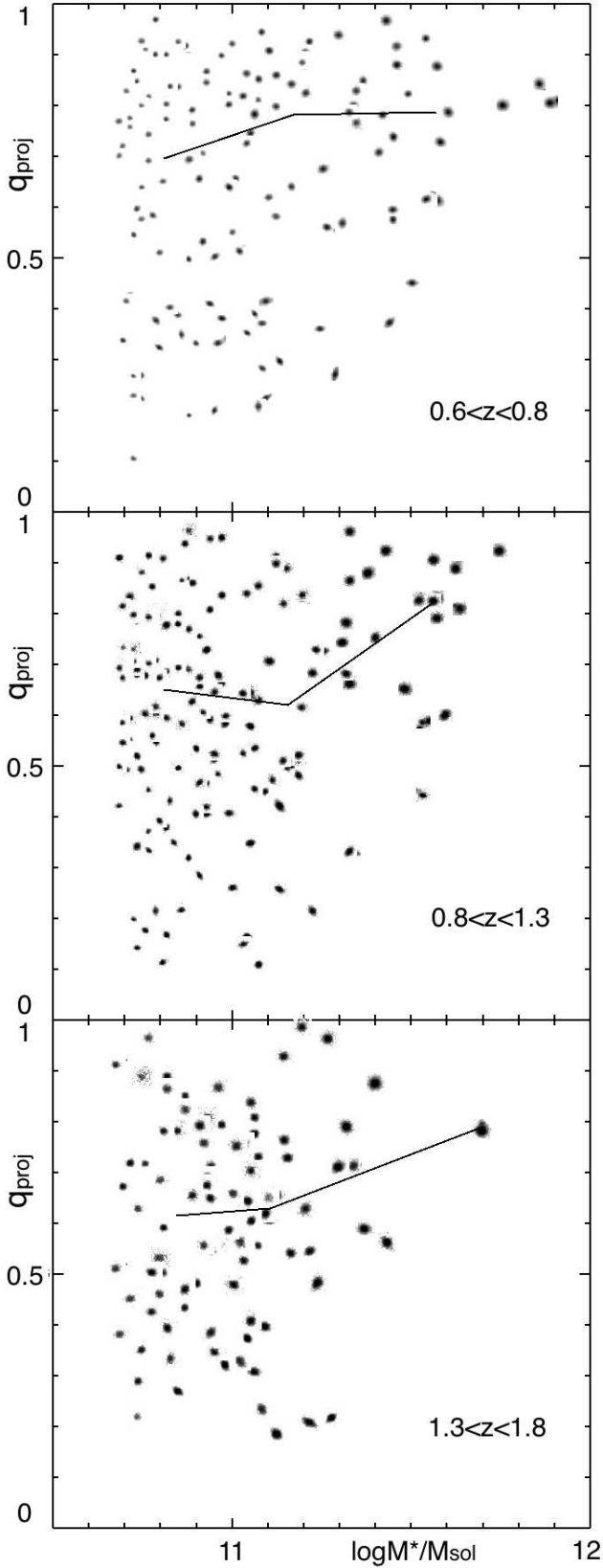}
\caption{The projected axis ratio versus stellar mass for our early- type galaxy sample, split into three redshift bins. The symbols are the HAWK-I $K$-band images of the galaxies. The lines represent running median values of the axis ratio. Up to our highest redshift bin, the most massive galaxies appear to be the roundest (see also Figure~\ref{mq_q} and Sec~3.1). Overall, the galaxies in the highest redshift bins are also flatter than their lower-redshift counterparts (see also Figure~\ref{cumulative_q} and Sec~3.2).}
\label{mq}
\end{figure}

In Figure~\ref{mq} we show the projected axis ratios of our sample of early-type galaxies as a function of their stellar mass, split into three redshift bins, each with $\sim100$ galaxies. 

Qualitatively speaking, Figure~\ref{mq} shows the same trend in all redshift bins: there appears to be a mass dependence for the projected axis-ratio, such as was seen in vdW09. As previously demonstrated by H12, at $z\sim0.7$ we find that the most massive ($log(M/M_\odot)>11.3$) early-type galaxies are round in projection and we find a lack of objects that are flat in projection among the most massive early-type galaxies. In Figure~\ref{mq} a similar trend is seen for early-type galaxies at $z>1$: at and below masses of $log(M/M_\odot)\sim11$ the galaxies in our sample show a broad range in axis ratios, whereas more massive galaxies are predominantly round. This is consistent with the conclusion reached by \citet{2011MNRAS.412..295T}, who found that very luminous radio galaxies at $z\sim 2$ have morphological properties similar to today's most massive elliptical galaxies in clusters.

\begin{figure}
\centering
\includegraphics[width=1.0\columnwidth]{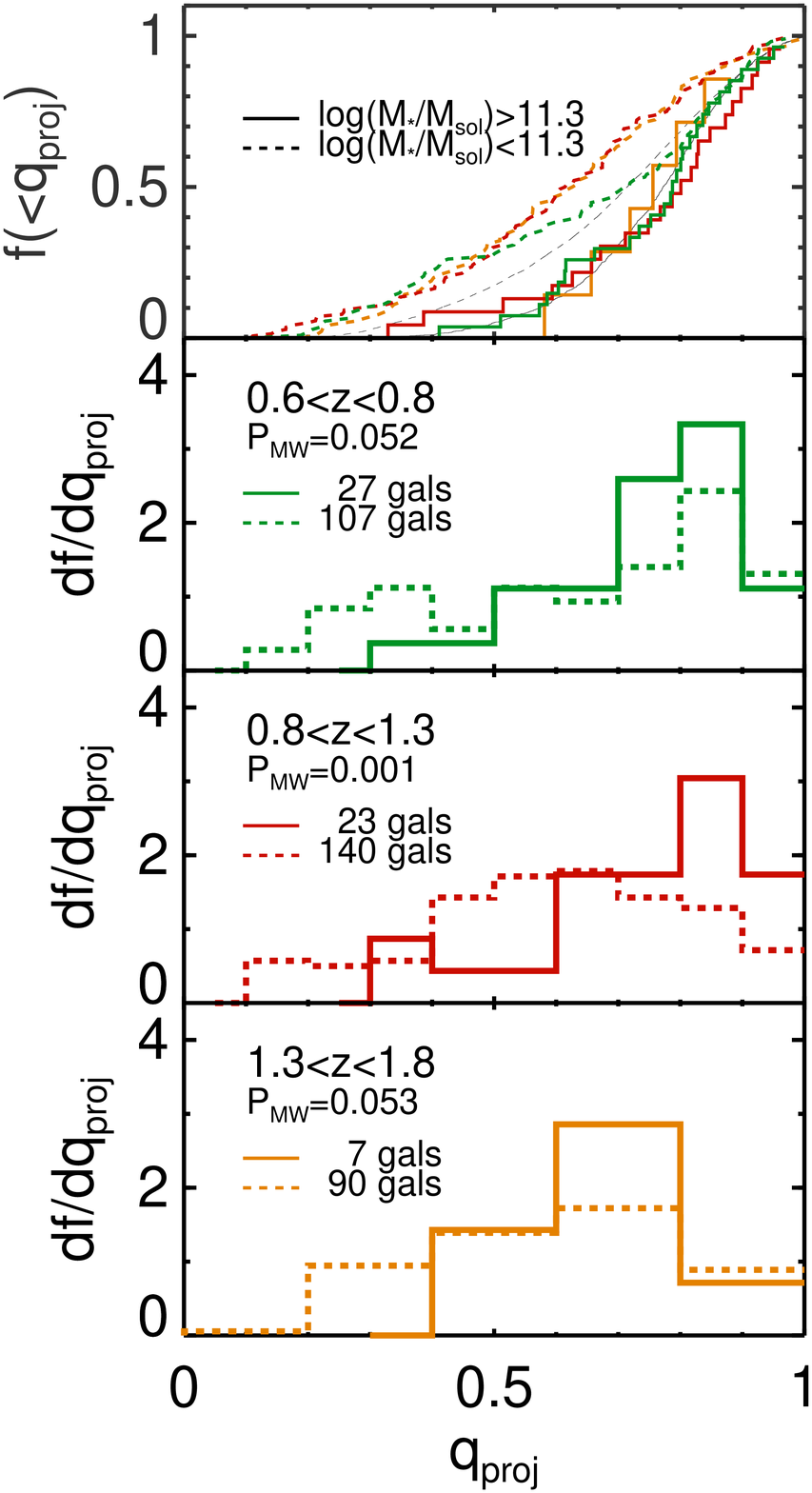}
\caption{Axis ratio histograms of early-type galaxies with masses $10.7<log(M/M_\odot)<11.3$ (dashed lines) and early-type galaxies with masses $log(M/M_\odot)>11.3$ (solid lines). The bottom panels show the distributions ($df/dq_{proj}$) of different redshift bins, as indicated; the top panel combine all redshift bins and show the cumulative histograms as a function of projected axis ratio ($f(<q_{proj}$), where the color coding corresponds to redshift, following the color coding in the bottom panels.  The top panel also shows in black the SDSS-based sample of local early-type galaxies from H12 in the same mass range. At all redshifts the most massive galaxies galaxies are the roundest; the significance of this observation is confirmed by the M-W statistical test (as indicated by the listed probabilities that the samples are statistically the same).}
\label{mq_q}
\end{figure}

In Figure~\ref{mq_q} we compare the axis-ratio distributions of galaxies with masses below and above $log(M/M_\odot)\sim11.3$. The Mann-Whitney U test shows that the trend seen in Figure~\ref{mq} is marginally significant in each of the redshift bins. For the combined sample of galaxies at redshifts $0.8 < z < 1.8$ we find that the most massive galaxies are rounder than the less massive galaxies with high confidence ($P=2.51\times10^{-4}$, i.e., a 3.66$\sigma$ result). For comparison, the equivalent value from the classifical Kolmogorov-Smirnov test is also low ($P_{KS}=1.81\times10^{-3}$, i.e., a 3.12$\sigma$ result).

Combining the results from vdW09, H12 and from this paper, we conclude that at all redshifts $0<z<2$ the most massive early-type galaxies are predominantly round in projection.  This implies that at all these cosmological epochs the formation mechanism for such massive galaxies precludes the formation of or requires the destruction of pre-existing disks.  Our interpretation is that merging, accompanied with little dissipation and star formation, has been the dominant formation channel for galaxies more massive than $log(M/M_\odot)=11.3$ since $z\sim2$. This process is reproduced in early-type galaxy formation models \citep[e.g.,][]{2009ApJ...699L.178N,2012ApJ...744...63O}, and the observed major merger rates up to $z\sim2$ are consistent with theoretical expectations \citep[e.g.,][]{2012ApJ...744...85M}.

\subsection{Shape Evolution at $1<z<2$ }

\begin{figure}
\centering
\includegraphics[width=1.0\columnwidth]{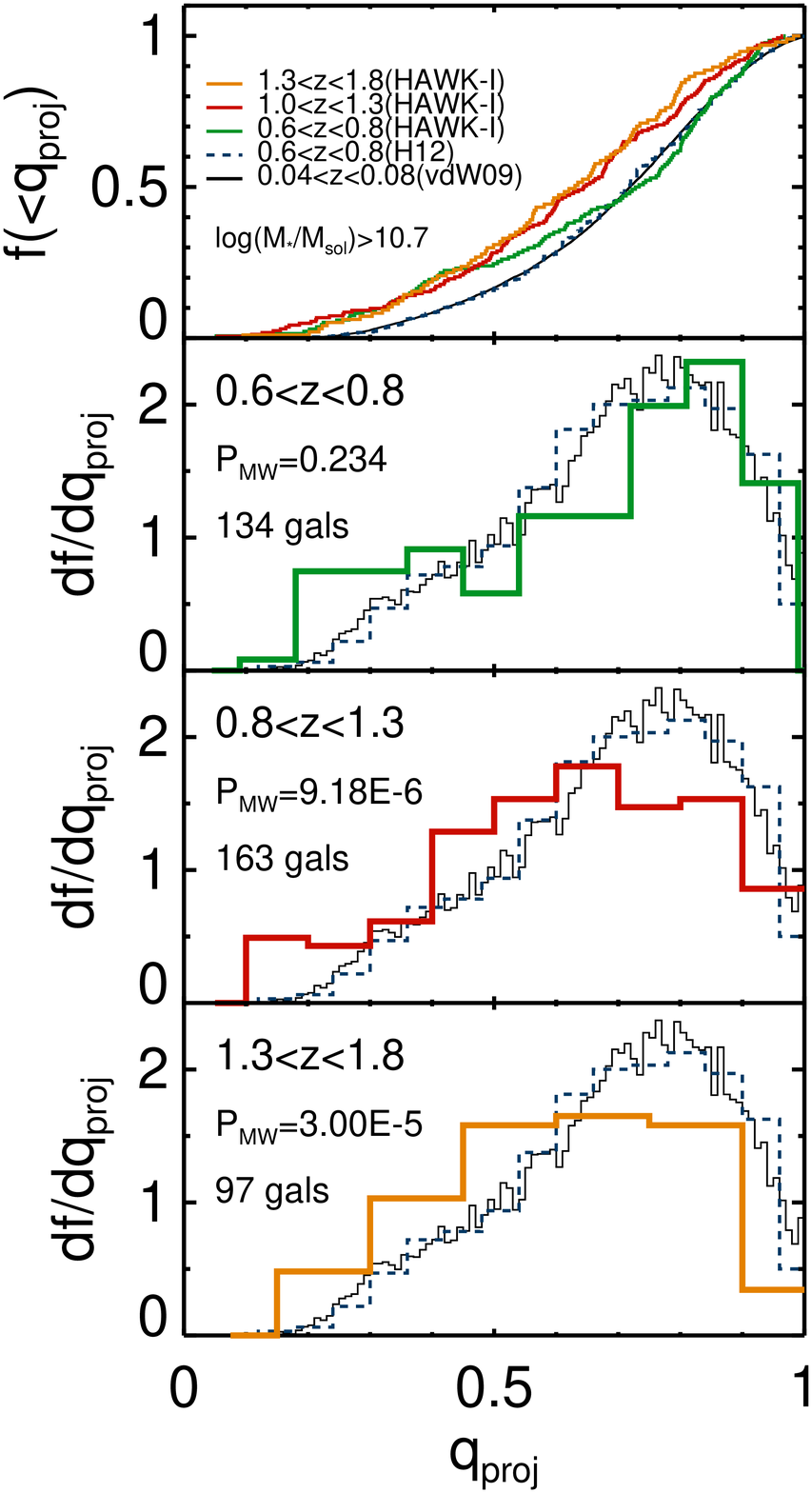}
\caption{Axis ratio histograms ($df/dq_{proj}$) of early-type galaxies with all masses $log(M/M_\odot)>10.7$ (our full sample) as a function of redshift in bottom three panels, always compared with the present-day sample from H12. The top panel shows cumulative histograms for all redshift bins as a function of projected axis ratio $f(<q_{proj}$). Up to z=0.8, as confirmed by the M-W statistical test (of which the probability is given that the samples are statistically the same) there is no significant redshift evolution in the projected axis-ratio distribution, consistent with the results from H12. At $z>1$, we find that early-type galaxies are flatter than
their present-day counterparts.}
\label{cumulative_q}
\end{figure}

Now we turn to the question whether the structural population properties evolve with redshift. vdW11 found an indication of a high incidence of disk-like early-type galaxies at $z\sim2$, but their sample is too small to confirm or rule out evolution with respect to the structure of early-type galaxies at the present day. In Figure~\ref{cumulative_q} we compare the shape distributions of early-type galaxies with masses $log(M/M_\odot)>10.7$ across a broad range in redshift, $0<z<1.8$. We reproduce the lack of strong evolution in the range $0<z<0.8$, reported by H12. 

\citet{2011MNRAS.413..921V} showed that cluster early-type galaxies
are \emph{rounder} at z=0.5-1 than in the local universe.  This result
is not necessarily at odds with our measurements: their signal is
mostly driven by galaxies below our mass limit and, moreover,
structural differences between cluster and field galaxies \citep[e.g.,
][]{2010ApJ...714.1779V} can be explained by environmental processes
such gas stripping that can produce flat early-type galaxies in
clusters.  An indication that the chosen mass range is relevant is
that low-mass galaxies show stronger environmental dependencies in
their properties than high-mass galaxies \citep[see,
e.g.,][]{2010ApJ...721..193P}.  Observations over a broader range of
environments and down to lower stellar masses than what is possible
with our data set are required to address these issues.

\begin{figure*}
\centering
\includegraphics[width=0.8\textwidth]{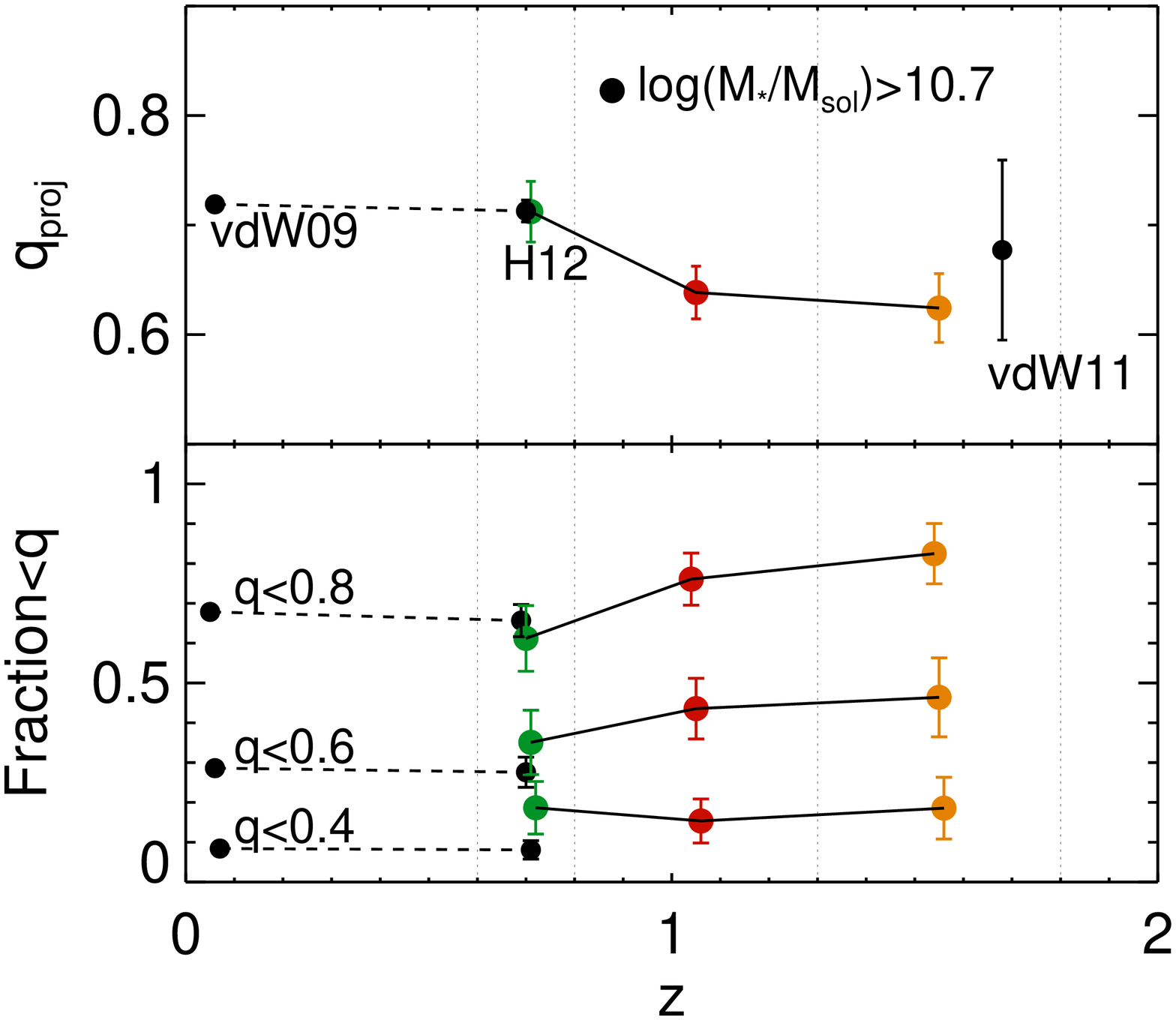}
\caption{Top panel: median axis ratio of early-type galaxies more massive than $log(M/M_\odot)>10.7$ as a function of redshift. The sample presented in this paper is represented by the colored data points; the black data points are taken from the literature as indicated. The error bars are estimated by bootstrapping. Bottom panel: fraction of galaxies with axis ratios smaller than indicated, again as a function of redshift. Beyond $z\sim1$, early-type galaxies typically have smaller axis ratios than present-day counterparts in the same mass range, indicative of a more disk-like structure. All error bars are inferred by 95\% confidence level compare to the same fraction.}
\label{q_z_distribution}
\end{figure*}

\begin{deluxetable*}{c|cccccc}
\tablewidth{0pc}
\tabletypesize{\scriptsize}
\tablecaption{Kolmogorov-Smirnov test: significant level ($log (M/M_\odot) > 10.7$)} 
\tablehead{
\colhead{$log(M/M_\odot)>10.7$} & \colhead{$1.53<z<2.31$} & \colhead{$ 1.3<z<1.8$} & \colhead{$0.8<z<1.3$} & \colhead{$0.6<z<0.8$} & \colhead{$0.6<z<0.8$} & \colhead{$0.04<z<0.08$}
}
\startdata
$1.53<z<2.31$ (vdW11) &  - & 0.463 & 0.549 & 0.671 & 0.668 & 0.688 \\
$1.3<z<1.8$ & 0.463 & - & 0.875 & *1.02E-2*\tablenotemark[a] & *2.58E-3* & *1.54E-4* \\
$0.8<z<1.3$ & 0.549 & 0.875 & - & *1.26E-2* & *8.93E-4* & *1.21E-4* \\
$0.6<z<0.8$ & 0.671 & *1.02E-2* & *1.26E-2* & - & 0.060 & 0.035 \\
$0.6<z<0.8$ (H12) & 0.668 & *2.58E-3* & *8.93E-4* & 0.060 & - & 0.584 \\
$0.04<z<0.08$ (vdW09) & 0.688 & *1.54E-3* & *1.21E-4* & 0.035 & 0.584 & - \\
\enddata
\tablenotetext{a}{Star symbol (*) represent the significant probability is smaller than 5\%. It implies the distributions are distinguishable.}
\label{ks}
\end{deluxetable*}

\begin{deluxetable*}{c|cccccc}
\tablewidth{0pc}
\tabletypesize{\scriptsize}
\tablecaption{Mann-Whitney test: significant level ($log (M/M_\odot) > 10.7$)} 
\tablehead{
\colhead{$log(M/M_\odot)>10.7$} & \colhead{$1.53<z<2.31$} & \colhead{$ 1.3<z<1.8$} & \colhead{$0.8<z<1.3$} & \colhead{$0.6<z<0.8$} & \colhead{$0.6<z<0.8$} & \colhead{$0.04<z<0.08$}
}
\startdata
$1.53<z<2.31$ (vdW11) &  - & 0.139 & 0.167 & 0.456 & 0.372 & 0.373 \\
$1.3<z<1.8$ & 0.139 & - & 0.315 & *1.12E-2* \tablenotemark[a] & *1.20E-4* & *3.00E-5* \\
$0.8<z<1.3$ & 0.167 & 0.315 & - & *2.06E-2* & *8.71E-5* & *9.18E-6* \\
$0.6<z<0.8$ & 0.456 & *1.12E-2* & *2.06E-2* & - & 0.249 & 0.234 \\
$0.6<z<0.8$ (H12)  & 0.372 & *1.20E-4* & *8.71E-5* & 0.249 & - & 0.445 \\
$0.04<z<0.08$ (vdW09) & 0.381 & *3.00E-5* & *9.18E-6* & 0.234 & 0.445 & - \\
\enddata
\tablenotetext{a}{Star symbol (*) represent the significant probability is smaller than 5\%. It implies the distributions are distinguishable.}
\label{mw}
\end{deluxetable*}

However, in our higher redshift bins ($z>0.8$) we see an excess of flat galaxies and a detectable evolution compared to the $z<0.8$ samples. A Kolmogorov-Smirnov (K-S) and Mann-Whitney U (M-W) tests shows that this trend is highly significant (see Tables~\ref{ks} and~\ref{mw}). Note that the axis ratio distribution from vdW11, due to their small sample size, is consistent with both the low- and high-z axis ratio distributions we analyze here.

Figure \ref{q_z_distribution} summarizes the observed evolution in the mean projected shape of early type galaxies.  Beyond $z\sim 1$ we see a gradual decrease in the median axis ratio of early-type galaxies, which is the result of an increased fraction of galaxies with axis ratios smaller than 0.6. The observed axis-ratio distributions imply that the intrinsic thickness of the typical early-type galaxy is no more than about 0.4. In a forthcoming paper we will quantify this intrinsic shape through detailed modeling.

The observations presented here can now put on a firmer footing the claim by vdW11 that many high-redshift early-type galaxies display a disk-like structure and that, plausibly, these galaxies will grow into larger, intrinsically round galaxies through merging.  The vdW11 sample had been too small to confirm or rule evolution at a fixed stellar mass, which is important in order to differentiate evolutionary paths that galaxies of a given mass take at different cosmic times.  Even at the present-day, the typical early-type galaxy with a mass below $log(M/M_\odot)=11$ is rather flattened (vdW09), which also corresponds to a disk-like kinematic structure (rotation) as shown by, for example, \citet{2011MNRAS.414..888E}.  Here we show that such galaxies had yet thinner intrinsic shapes at $z>1$.  Good correspondence between rotation of the stellar body and flattening has been confirmed up to $z\sim 1$ by \citet{2008ApJ...684..260V}.  Although such confirmation is, so far, lacking at higher redshifts it is reasonable to assume that flattening implies rotation at all redshifts.

We note that beyond the simple observation that early-type galaxies,
here defined as galaxies with little star-formation activity, are on
average flatter at $z>1$ than at lower redshifts, we cannot
distinguish between the different varieties of such galaxies -- such
as quiescent spirals, barred/ringed S0s, etc. -- and their evolution.

Our results do not argue against a morphological transition that occurs along with the truncation of star formation, as argued by, for example, \citet{2012ApJ...753..167B} on the basis of the high S\'{e}rsic indices of quiescent galaxies: high-redshift early-type galaxies such as those studied here, despite the generally disk-like character inferred from their axis ratio distribution, are not pure disks and have higher S\'{e}rsic indices than star forming galaxies \citep[e.g.,][]{2011ApJ...742...96W}. Based on bulge-disk decompositions, \citet{2012arXiv1206.4322B} arrive at the
conclusion that the early-type galaxy population at $z>1$ is mix of
bulge- and disk-dominated galaxies, indicating that the transition
from actively star-forming to quiescent need not always coincide with
the formation of a dominant bulge.
Taken together, these observations are consistent with a picture in which gas had time to settle in a disk before star formation was truncated, and that this disk wholly or partially survived the process that truncated star formation.

\section{Conclusions}

We measured the projected axis ratios from VLT/HAWK-I $K$-band imaging of a sample of early-type galaxies in the redshift range $0.6<z<1.8$ selected by their low specific star formation rates.  We find that at all redshifts $z<2$ the most massive galaxies $log(M/M_\odot)>11.3$ are predominantly round. The lack of very massive, highly flattened galaxies suggests a universal ceiling mass for the formation of disks, independent of cosmic epoch.  In order for galaxies to grow beyond this ceiling mass, separate evolutionary channel, presumably merging, has to be invoked.

In the full sample ($log(M/M_\odot)>10.7$), we find, at all redshifts, a large range in projected axis ratios, reflecting a more disk-like structure.  In addition, we find quantitative evidence that early-type galaxies at $z>1$ are more disk-like than their equally massive, present-day counterparts.  Therefore, for most early-type galaxies gas had time to settle into a disk before star formation ceased to produce a more passive galaxy.  Plausibly, such galaxies grow in mass over time, through mostly dissipationless merging and accretion of satellites, losing some of the angular momentum, and growing in size.

\acknowledgments
Y.-Y. Chang was funded by the IMPRS for Astronomy \& Cosmic Physics at the University of Heidelberg and the Marie Curie Initial Training Network ELIXIR of the European Commission under contract PITN-GA-2008-214227.

\end{document}